# دراسة أداء خوارزمية CEEMDAN في فصل المنابع الصوتية في بيئات منخفضة التحديد

رواد ملحم[1]، رياض حمادة[2]، آصف جعفر[3]

[1] طالب دراسات عليا، ماجستير في هندسة نظم الاتصالات، قسم الاتصالات، المعهد العالي للعلوم التطبيقية والتكنولوجيا، rawad.melhem@hiast.edu.sy

[2] أستاذ في قسم الاتصالات، المعهد العالي للعلوم التطبيقية والتكنولوجيا.

[3] أستاذ في قسم الاتصالات، المعهد العالي للعلوم التطبيقية والتكنولوجيا.

المعهد العالي للعلوم التطبيقية والتكنولوجيا

HIAST

## الملخص:

تعتبر خوارزمية Complete Ensemble Empirical Mode Decomposition with Adaptive Noise (CEEMDAN) من الطرق الحديثة المستخدمة في تحليل الإشارات غير المستقرة. يقدم البحث دراسة لفعّالية هذه الطريقة في فصل المنابع الصوتية والكلامية من خلال دراسة تجريبية تسمح بمعرفة حدود عملها، وأُستنتج شرطان متعلقان بترددات ومطالات الإشارتين الممزوجتين لتتمكن الخوارزمية من فصلهما، ثم دُرس أداء الخوارزمية في فصل الضجيج عن الكلام وفصل الإشارات الكلامية عن بعضها، وتوصّل البحث إلى نتيجة تفيد بقدرة الطريقة على فصل بعض أنواع الضجيج عن الكلام (تحسين الكلام)، وعجزها عن فصل الإشارات الكلامية عن بعضها. تمّت المحاكاة باستخدام بيئة الماتلاب وقاعدة المعطيات Noizeus.

**الكلمات المفتاحية**: خوارزمية CEEMDAN، فصل المنابع الصوتية، تحسين الكلام، بيئات منخفضة التحديد.

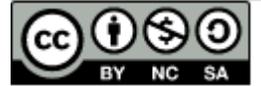

# Study of the Performance of CEEMDAN in Underdetermined Speech Separation

**Rawad Melhem[1] , Riad Hamadeh[2] , Assef Jafar[3]**

[1]Postgraduate Student, Department of Telecommunications Engineering, Higher Institute for Applied Sciences and Technology,  rawad.melhem@hiast.edu.sy

[2]Professor, Department of Telecommunications Engineering, Higher Institute for Applied Sciences and Technology.

[3]Professor, Department of Telecommunications Engineering, Higher Institute for Applied Sciences and Technology.

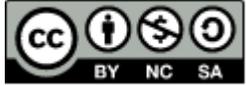

**Abstract**

The CEEMDAN algorithm is one of the modern methods used in the analysis of non-stationary signals. This research presents a study of the effectiveness of this method in audio source separation to know the limits of its work. It concluded two conditions related to frequencies and amplitudes of mixed signals to be separated by CEEMDAN. The performance of the algorithm in separating noise from speech and separating speech signals from each other is studied. The research reached a conclusion that CEEMDAN can remove some types of noise from speech (speech improvement), and it cannot separate speech signals from each other (cocktail party). Simulation is done using Matlab environment and Noizeus database.

 **Keyword:** CEEMDAN Algorithm, Speech Separation, Speech Enhancement Underdetermined Separation.

1- المقدمة

تبرز أهمية فصل المنابع الصوتية في العديد من التطبيقات مثل تحويل الكلام لنص مكتوب Automatic meeting transcription، وتسمية المقاطع الصوتية بشكل آلي automatic captioning for audio/video recordings مثل موقع youtube، والتطبيقات التي تحتاج لتفاعل الانسان مع الآلة عن طريق الصوت كما في تجهيزات انترنت الأشياء IOT، والأجهزة المساعدة للسمع Hearing aids.

تقسم مسألة فصل المنابع الصوتية إلى نوعين الأولى منخفضة التحديد Underdetermined حيث يكون عدد الميكروفونات أقل من عدد المنابع الصوتية وهي الحالة الأصعب، والثانية عالية التحديد Overdetermined حيث يكون عدد الميكروفونات أكثر من عدد المنابع الصوتية. يعالج هذا البحث النوع الأول "منخفضة التحديد" حيث تم استخدام ميكروفون وحيد لتسجيل إشارة مزيج لمنبعين صوتيين مختلفين، وتسمى أيضاً هذه الحالة فصل إشارتين بقناة وحيدة single-channel separation.
أُستخدمت تقنية طرح الطيف (Verteletskaya, et al., 2011) Spectral Subtraction (SS) لفصل الضجيج عن الكلام، حيث تمّ تقدير طيف الضجيج باستخدام تقنيات كشف الكلام Voice Activity Detection (VAD) ثم يُطرح من طيف الإشارة المضججة. كانت النتائج جيدة، ولكن تعاني طرق طرح الطيف من عدة عيوب منها ظهور الضجيج الموسيقي (وهو عبارة عن مركبات جيبية بترددات عشوائية تظهر وتختفي في أطر زمنية قصيرة من الإشارة)، وعدم ملاءمة هذه الطريقة في حالة الضجيج الملوّن colored noise حيث ينخفض أداؤها كثيراً بوجود هذا النوع من الضجيج مقارنة مع حالة الضجيج الأبيض، لأن الضجيج الملوّن يظهر في حزم ترددية معينة باستطاعة عالية وتنخفض استطاعته كثيراً في باقي الحزم مما يستدعي معالجة أصعب لحذفه، ليس كالضجيج الأبيض الذي يمتد على كامل المجال الترددي بنفس الاستطاعة. أُضيف لطرق طرح الطيف مرشح وينر (Gomez, et al., 2010) لحذف الضجيج الملوّن، حيث يُقَدَّر الضجيج في كل إطار (بعد تقسيم الإشارة إلى أطر متراكبة)، وتعدَّل معاملات المرشح عند كل إطار، وهذا ما يجعله قادراً على التعامل مع الضجيج الملون. يؤدي الخطأ أثناء تقدير الضجيج في كل إطار إلى انخفاض أداء مرشح وينر، وهو أهم مشاكل هذا المرشح، وبالتالي يفيد استخدام مرشح وينر مع Minimum Mean Square Error (MMSE) في تحسين النتائج (Krawczyk-Becker, et al., 2015)، وعموماً فإن نتائج طرق تحسين الكلام التي تعتمد على الخصائص الإحصائية أكثر دقةً، وتُستخدم في الكثير من التطبيقات العملية ولكن تبقى مسألة تقدير معاملات النموذج الإحصائي غير دقيقة وتتأثّر بقيم نسبة الإشارة للضجيج SNR الصغيرة، ومازالت موضع جدل ونقاش. تمكن (So, et al., 2017) من فصل الضجيج الملوّن عن الكلام باستخدام طرق الترشيح المتكيف Adaptive Filtering Methods، ومن أهم هذه المرشحات مرشح كالمان Kalman Filter والذي يُجري تقدير أمثلي لمعاملات إحصائية للنظام، ولكن على حساب التعقيد الحسابي والكيان الصلب للنظام Hardware.

اعتمد الباحثان Krishna و Ramaswamy في (Prasanna, et al., 2017) خوارزمية EMD مع تحويل هوانغ هيلبرت Huang Hilbert Transform (HHT) لفصل إشارتين كلاميتين عن بعضهما، حيث تمّ الحصول على الترددات والمطالات اللحظية لإشارة المزيج ثم إجراء تصنيف حسب الترددات والمطالات لفصل الإشارتين الكلاميتين، ولكن لدى تنجيز هذه الخوارزمية خلال هذا البحث باستخدام الماتلاب لم يتم الفصل، ويعود ذلك لكون التصنيف حسب الترددات والمطالات ليس صحيحاً بسبب التداخل الكبير في ترددات

ومطالات الإشارات الكلامية في المزيج، بالإضافة إلى أن مصطلح الترددات اللحظية متناقض ويسبب التباس وهو غير صحيح دائماً(Norden, et al., 1998) .

عرضت الورقة (Po-Sen, et al., 2015) استخدام الشبكات العصبونية لفصل الاشارات الكلامية، وكانت نتائج الفصل دقيقة، وتعاني هذه الطريقة من مشكلة تبديل المنابع وعدد مخارج الشبكة الثابت (كل مخرج للشبكة يمثل إشارة منفصلة). عُولجت هذه المشاكل في أبحاث أخرى (Yu, et al., 2017) (Zhuo, et al., 2017)، ورغم ذلك فهي غير مرنة في الفصل، وتحتاج لزمن طويل للتدريب.

جرى استخدام شبكة عصبونية لحذف الضجيج من الكلام في (Pascual Santiago, et al., 2017)، بالاعتماد على شبكة عصبونية من النوع Generative Adversarial Networks (GAN)، وأُطلق على هذه الطريقة Speech Enhancement GAN (SEGAN)، وأعطت SEGAN نتائج جيدة.

يهتم هذا البحث بشكل أساسي بتقييم خوارزمية CEEMDAN (María, et al., 2011) وبيان حدود عملها. وهي نسخة مطوَّرة عن الخوارزمية Empirical Mode Decomposition(EMD) التي ظهرت في التسعينيات من القرن الماضي، وهي تحلل الإشارة إلى مجموعة أنماط يمكن أن يكون كل نمط ناتجاً عن منبع معين، وجرى استخدامها لفصل العديد من الاشارات عن بعضها، مثل فصل الإشارات الحيوية المأخوذة من الحساسات الطبية (Manuel, et al., 2008)، وفصل الإشارات الملتقطة من أجهزة التنبؤ بالزلازل (Jiajun & Baan, 2015)، كما أُستخدمت مؤخراً لفصل الضجيج عن الكلام ولكن لم تعطِ نتائج جيدة لوحدها (Taufiq & Hasan, 2009)، لذلك تمّ مكاملتها مع أنظمة أخرى مثل مرشح وينر وغيره. عانت خوارزمية EMD من مشكلة مزج الأنماط Mode Mixing فقد يظهر بعد تحليل الإشارة نمطين متشابهين كثيراً في التردد، أو نمطين ممزوجين في نمط واحد. عُولجت هذه المشكلة بإضافة سلاسل من الضجيج الأبيض الغوصي للإشارة الأصلية، وفي كل مرة تُطبَّق EMD على الإشارة المضاف إليها سلسلة ضجيج وتُحسب الأنماط، وبعد ذلك يتم حساب الوسطي للأنماط المتوافقة الناتجة عن إعادة تطبيق EMD، سُميت هذه الطريقة Ensemble Empirical Mode Decomposition (EEMD) (Zhaohua & Huang, 2009)، حيث سمحت إضافة سلاسل الضجيج الأبيض بالاستفادة من سلوك EMD المشابه لبنك مرشحات على كامل المجال الطيفي (Zhaohua & Huang, 2009). قدَّمت خوارزمية EEMD حلاً ناجعاً لمشكلة مزج الأنماط، ولكنها أضافت مشكلة أخرى وهي عدم الحصول على الإشارة الأصلية بدقة عند جمع الأنماط بسبب سلاسل الضجيج الأبيض المضافة. تمّ تطوير خوارزمية CEEMDAN لحل هذه المشكلة، حيث استخدمت سلاسل ضجيج بطريقة مختلفة فأصبح خطأ إعادة بناء الإشارة معدوماً، وزادت دقة فصل الأنماط، وقلَّ التعقيد الحسابي (María, et al., 2011).

رغم تطبيق خوارزمية EMD مؤخراً في فصل المنابع الصوتية، إلا أن خوارزمية CEEMDAN لم تُستخدم لفصل المنابع الصوتية حتى الآن، ومن هنا يبرز أهمية هذا البحث في تحديد جدوى هذه الخوارزمية في هذا المجال. نظراً لكون مصطلح فصل المنابع الصوتية يغطي العديد من الحالات مثل فصل الضجيج عن الكلام (تحسين الكلام)، أو فصل صوت آلة موسيقية عن أخرى، أو فصل اشارة كلامية عن أخرى، فقد اهتمّ هذا البحث بدراسةَ أداء خوارزمية CEEMDAN في فصل الضجيج عن الكلام وفي فصل إشارتين كلاميتين عن بعضهما. تقوم الخوارزمية التكيفية CEEMDAN بتحليل الإشارة إلى مجموعة أنماط، قد يكون كل نمط هو منبع صوتي مستقل، ستقدم هذه الورقة شرحاً عن إمكانياتها وحدود عملها بالاعتماد

على عدة معايير للتقييم (Vincent, et al., 2006) أهمها نسبة الاشارة للتشوه SDR، ونسبة الإشارة للضجيج الصنعي SAR، ومعيار وضوح الإشارة Perceptual Evaluation Speech Quality (PESQ).

يُوصَّف التشوه الحاصل للإشارة المحسَّنة بالعلاقة التالية:

$$D_{total} = \frac{\| s_n - \widehat{s_n} \|^2}{\| s_n \|^2} \qquad (1)$$

يعبر $\|.\|^2$ عن طاقة الإشارة.
$s_n$ الإشارة الأصلية، $\widehat{s_n}$ الإشارة المحسَّنة.
وتكون نسبة الإشارة للتشوه هي:

$$SDR = 10 \, log_{10} D_{total}^{-1} \qquad (2)$$

يُعطى التشوه الناتج عن وجود ضجيج صنعي بالعلاقة:

$$D_{artif} = \frac{\| e_{artif} \|^2}{\| \langle \widehat{s_n}, s_n \rangle s_n + e_{interf} + e_{noise} \|^2} \qquad (3)$$

حيث $e_{interf}, e_{noise}, e_{artif}$ الضجيج الصنعي والجمعي وضجيج التداخل على الترتيب.

وتكون نسبة الإشارة للتشوه الصنعي هي:

$$SAR = 10 log_{10}(D_{artif}^{-1}) \qquad (4)$$

تقدم الفقرة (2) شرحاً لخوارزمية CEEMDAN، ويُركِّز في الفقرة (3) على إيجابيات وسلبيات خوارزمية EMD (التي هي نواة CEEMDAN) مع بيان مفصل لمشكلة مزج الأنماط التي تعاني منها خوارزمية EMD. تعرض الفقرة (4) نتائج الدراسة التجريبية التي تمّت في إطار هذا البحث لمعرفة سلوك CEEMDAN في فصل الإشارات الممزوجة وتُستنبط شروط عملها. تُظهر الفقرة (5) إمكانية CEEMDAN في فصل المنابع الصوتية مع عرض النتائج التي حصلنا عليها.

## 2- خوارزمية Complete Ensemble Empirical Mode Decomposition with Adaptive Noise (CEEMDAN)

تعتبر خوارزمية EMD خوارزمية تكرارية Iterative Algorithm لتحليل الاشارات غير المستقرة إلى توابع تسمى توابع الأنماط الأساسية(IMF) Intrinsic Mode Function.

يُشترط أن يحقق تابع النمط الأساسي ما يلي (Norden, et al., 1998):

1- أن يتساوى عدد النهايات المحلية الكبرى والصغرى لمنحني التابع مع عدد التقاطعات الصفرية أو يختلف عنها بواحد.

2- أن تكون القيمة الوسطى للغلاف العلوي والغلاف السفلي في أي نقطة من التابع معدومة.

إن تابع IMF هو تابع وحيد المركبة monocomponent، مع العلم أنه لا يوجد تعريف رياضي لتابع وحيد المركبة [6]، ولكنه يعني أن الإشارة تحوي قيمة وحيدة للتردد لذلك سُمي وحيد المركبة.

تُعتبر الأنماط الأساسية التي يتم الحصول عليها أثناء تحليل الإشارة باستخدام خوارزمية EMD توابع قاعدة base functions للإشارة، وهي تقابل توابع الجيب والتجيب في تحويل فورييه ويمكن كتابة الإشارة y بدلالة المركبات الأساسية كما يلي:

$$y(n) = \sum_{k=1}^{M} C_k(n) + r_M(n) \qquad (5)$$

حيث $C_k$ تابع المركبة الأساسي رقم $k$، و r الباقي من عملية تحليل الإشارة y.

إن أهم ما يميز هذه الخوارزمية أنها لا تفترض قيود أو شروط على الإشارة المدروسة، فهي تعتمد على خصائص الإشارة

المحلية (نهايات عظمى وصغرى) وبالتالي يختلف عدد هذه المركبات حسب الإشارة المدروسة لذلك تعتبر طريقة متكيفة أو مقادة بالمعطيات data-driven method وهي مناسبة لتحليل الإشارات غير المستقرة أكثر من تحويل فورييه وويفلت (Norden, et al., 1998). ولما كانت الإشارة الكلامية غير مستقرة فقد يكون تحليل الإشارات الكلامية بهذه الخوارزمية مجدياً في تحسينها.

يبين الشكل (1) خطوات تنفيذ هذه الخوارزمية:

1- إيجاد النهايات المحلية العظمى والصغرى للإشارة.
2- إيجاد الغلاف العلوي والسفلي للإشارة بإجراء استيفاء للنهايات المحلية.
3- حساب وسطي الغلافين ثم طرحه من الإشارة الأصلية فنحصل على باق $m(t)$.
4- اختبار إن كان $m(t)$ هو تابع IMF أم لا، في حال "لا" تعاد الخطوات السابقة حتى نحصل على أول مركبة أساسية IMF.
5- عند الحصول على أول مركبة أساسية تُطرح من الإشارة الأصلية، وتُعاد الخطوات السابقة على ناتج الطرح (الباقي) للحصول على ثاني مركبة أساسية، وهكذا للوصول إلى باق لا يمكن تحليله.

يُحدد عدد التوابع الأساسية أثناء التحليل (شرط توقف تكرار الخوارزمية) بحساب الانحراف المعياري بين مركبتين متتاليتين، والذي يجب أن يكون أصغر من عتبة معينة. يعتبر معيار تقارب نمط كوشي معياراً جيداً لتحديد عدد التوابع الأساسية، ويعطى بالعلاقة:

$$SD = \sum_{n=0}^{N-1}\left[\frac{(C_{k-1}(n)-C_k(n))^2}{C_{k-1}^2(n)}\right] \quad (6)$$

حيث $C_k$ هو المركبة الأساسية رقم $k$.

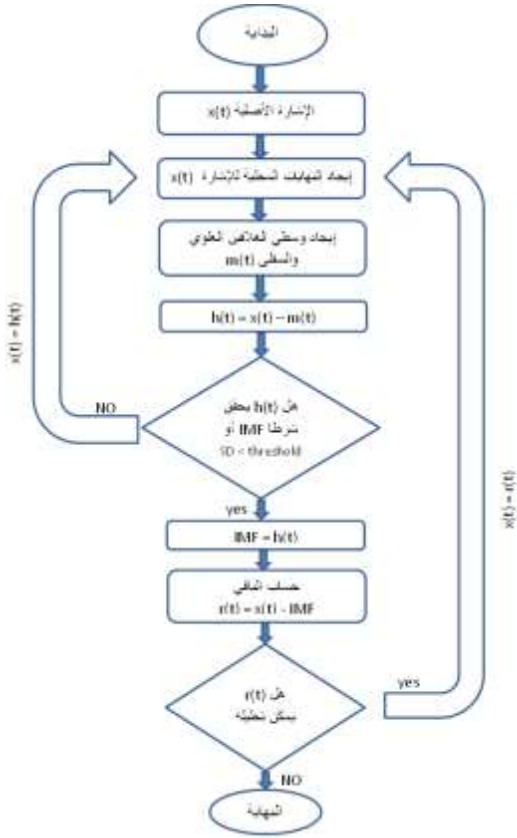

**الشكل-1- مخطط تدفقي لخوارزمية EMD**

### 3- سلبيات وإيجابيات خوارزمية EMD

*السلبيات:*

- لا تصلح خوارزمية EMD للعمل على نحو متدفق Streaming بسبب عدم القدرة على التنبؤ بعدد المركبات الأساسية، حيث يمكن لكل نافذة زمنية أن تعطي عدد مختلف من المركبات، وبالتالي لا يمكن جمعها مع النوافذ اللاحقة. في تحويلي فورييه وويفلت يكون عدد توابع القاعدة ثابت وبالتالي فهما مناسبان أكثر للعمل في التحليل المستمر Online Decomposition.
- مزج الانماط Mode Mixing (سيتم شرحه في الفقرة 3-1).

- عدم وجود برهان رياضي واضح أو أساس نظري لهذه الخوارزمية (Kopsinis, 2008).

*الإيجابيات:*

أهم ما يميز هذه الخوارزمية أنها لا تفترض شروط أو قيود على الإشارة فهي مقادة بالمعطيات Data-Driven، وتحلل الإشارات غير المستقرة.

### 3-1- مزج الأنماط Mode Mixing

تقوم فكرة خوارزمية EMD على اعتبار الإشارة المدروسة هي مزيج من أنماط IMFs. كل نمط هو تابع رياضي له خاصيات رياضية وله معنى فيزيائي (يمثل منبع ما)، عندما يتم تداخل بعض الأنماط تفقد تلك المركبات المعنى الفيزيائي الخاص بها. يحدث مزج الأنماط من عدة عوامل أهمها (B & Molinas, 2017):

- وجود مركبات ترددية متجاورة في المزيج تنتمي لمنابع مختلفة.
- التقطّع في الإشارة (مثل الإشارة الكلامية).
- الضجيج.

لحل مشكلة مزج الأنماط تمّ إضافة سلاسل ضجيج للإشارة أثناء التحليل، ثم حساب الوسطي لها (Zhaohua & Huang, 2009)، سُميت هذه الطريقة Ensemble Empirical Mode Decomposition (EEMD).

### 3-2- خوارزمية EEMD

تُمزج في هذه الطريقة الإشارة المدروسة مع عدد من سلاسل ضجيج أبيض غوصي، ويُحلل كل مزيج باستخدام EMD، ثم يؤخذ وسطي الأنماط المتوافقة كما في الخطوات التالية:

1- مزج الإشارة المدروسة مع عدد I من سلاسل الضجيج

$$x^i[n] = x[n] + w^i[n] \quad (7)$$

حيث $w^i$ سلسلة الضجيج رقم i وعدد هذه السلاسل هو I.

$x$ الإشارة المدروسة.

2- يتم تحليل كل $x^i[n]$ بخوارزمية EMD، فنحصل على عدد K من الأنماط $IMF_k^i$

3- يتم حساب الوسطي للأنماط المتوافقة

$$\overline{IMF_k}[n] = \frac{1}{I}\sum_{i=1}^{I} IMF_k^i[n] \quad (8)$$

يؤدي تحليل كل $x^i[n]$ بشكل مستقل عن التحقيقات الأخرى للضجيج إلى حدوث خطأ في إعادة بناء الإشارة. تمّ تدارك هذا الخطأ في خوارزمية CEEMDAN.

### 3-3- خوارزمية CEEMDAN

يُرمز للأنماط الناتجة عن التحليل بهذه الخوارزمية بالرمز $\widetilde{IMF}_k$ وذلك لتمييزها عن الأنماط الناتجة باستخدام الخوارزميات السابقة. يتم حساب النمط الأول تماماً كما في الخوارزمية EEMD، ويكون الباقي

$$r_1[n] = x[n] - \widetilde{IMF}_1[n] \quad (9)$$

يُحسب النمط الثاني بالعلاقة:

$$\widetilde{IMF}_2[n] = \frac{1}{I}\sum_{i=1}^{I} E_1\left(r_1[n] + \varepsilon_1 E_1(w^i[n])\right) \quad (10)$$

حيث يدل $E_1(.)$ على النمط الأول الناتج من تطبيق EMD على الإشارة.

$\varepsilon_1$ هو معامل لتوزين مطال الضجيج، حيث يتم اختياره صغيراً للترددات العالية، وبالعكس للترددات المنخفضة (Zhaohua & Huang, 2009)

الباقي الثاني هو:

$$r_2[n] = r_1[n] - \widetilde{IMF}_2[n] \quad (11)$$

وهكذا تكون قد رُبطت عمليات التحليل ببعضها.

يتم حساب الباقي من أجل k>=2 بالعلاقة:

$$r_k[n] = r_{k-1}[n] - \widehat{IMF}_k[n] \quad (12)$$

ويُحسب كل نمط من أجل k>=2 بالعلاقة:

$$\widehat{IMF}_{k+1}[n] = \frac{1}{I}\sum_{i=1}^{I} E_1\big(r_k[n] + \varepsilon_k E_k(w^i[n])\big) \quad (13)$$

تُكرر الخطوات حتى نحصل على باقٍ لا يمكن تحليله بطريقة EMD (أي يحوي على الأكثر نهاية محلية عظمى أو صغرى).

### 4- دراسة قدرة وإمكانية خوارزمية CEEMDAN على فصل الإشارات المتداخلة

لدراسة قدرة خوارزمية CEEMDAN على فصل الإشارات الصوتية الممزوجة جرى في إطار هذا البحث القيام بتمثيل حاسوبي لهذه الخوارزمية باستخدام MATLAB وفق مايلي:

1- تمثيل حالة وجود إشارة المزيج عبارة عن مركبتين جيبيتين، ومحاولة فصلهما بالخوارزمية المذكورة تبعاً للاختلاف والتشابه في ترددي ومطالي الإشارتين.

2- تطبيق CEEMDAN على إشارات كلامية ممزوجة بضجيج تمّ الحصول عليها من قاعدة بيانات Noizeus لتحديد قدرة هذه الخوارزمية على فصل الضجيج عن الكلام.

3- تطبيق CEEMDAN على إشارة كلامية ممزوجة بأخرى لتحديد مقدرة الخوارزمية على فصل إشارتين كلاميتين.

اعتمدت نسبة الإشارة للتشوه SDR كمعيار لتقييم أداء الفصل بين الإشارات.

### 4-1- دراسة تأثير اختلاف الترددات على أداء الخوارزمية

تُجرى التجربة الأولى على حالة إشارتين جيبيتين ممزوجتين ترددهما F1 وF2 ولهما نفس المطال، بتطبق CEEMDAN على المجموع ومقارنة الأنماط الناتجة مع الإشارتين الأصليتين (بالاعتماد على المعيار SDR). تُكرر هذه التجربة مع تغيير F1 وF2 بحيث تبدأ بقيم متباعدة ثم تتقارب كما هو مبين في الجدول (1).

| F1(HZ) | F2(HZ) | SDR(dB) |
|---|---|---|
| 700 | 300 | 24.67 |
| 700 | 350 | 23.96 |
| 700 | 400 | 19.13 |
| 700 | 450 | 12.75 |
| 700 | 500 | 1.72 |
| 700 | 550 | −0.01 |
| 700 | 600 | −1.73 |
| 700 | 650 | −1.83 |
| 700 | 750 | −1.91 |
| 700 | 800 | −1.77 |
| 700 | 850 | −0.83 |
| 700 | 900 | −0.46 |
| 700 | 950 | 6.99 |
| 700 | 1000 | 9.69 |
| 700 | 1050 | 13.09 |
| 700 | 1100 | 15.82 |
| 700 | 1150 | 17.42 |
| 700 | 1200 | 19.29 |
| 700 | 1250 | 19.86 |
| 700 | 1300 | 21.65 |
| 700 | 1350 | 22.64 |

**الجدول (1) يبين قيم SDR عند تغيير تردد إحدى إشارتي المزيج**

يبين الشكل –2– قيمة SDR بدلالة نسبة الترددين حيث يتضح من الشكل تحسُّن الفصل كلما ابتعد الترددان عن بعضهما.

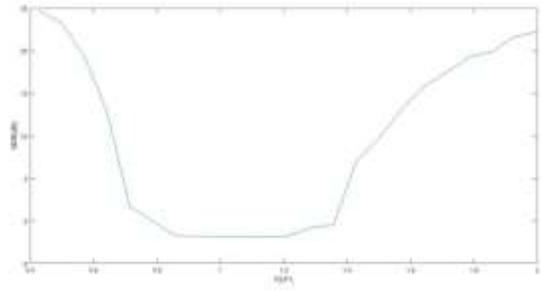

**الشكل-2- نسبة الاشارة للتشوه بدلالة نسبة ترددي الإشارتين الممزوجتين**

يُلاحظ من المنحني السابق أن نسبة الإشارة للتشوه SDR في مخرجي الخوارزمية يكون أكبر من 15dB إذا كان نسبة الترددين خارج المجال [0.6,1.6] أي بمعنى آخر تستطيع خوارزمية CEEMDAN أن تفصل إشارتين جيبيتين لهما نفس المطال على نحو جيد إذا كانت نسبة تردد إحداهما إلى الأخرى أكبر من 1.6 أو أقل من 0.6.

### 4-2- دراسة تأثير تشابه واختلاف المطالات على أداء الخوارزمية

بإعادة التجربة السابقة باستخدام مزيج من مركبتين جيبيتين حققتا أعلى نسبة فصل وهي F1=700Hz و F2=300Hz مع القيام بتغيير نسبة مطاليهما كما في الجدول (2) للحصول على مدى تأثير المطالات النسبية للإشارات الممزوجة على أداء الخوارزمية.

| A1 | A2 | SDR(dB) |
|---|---|---|
| 1 | 0.2 | −6.94 |
| 1 | 0.3 | −2.56 |
| 1 | 0.4 | 5.20 |
| 1 | 0.5 | 15.26 |
| 1 | 0.6 | 20.98 |
| 1 | 0.7 | 21.20 |
| 1 | 0.8 | 23.42 |
| 1 | 0.9 | 23.42 |
| 1 | 1.0 | 24.31 |
| 1 | 1.2 | 22.67 |
| 1 | 1.5 | 22.04 |
| 1 | 1.9 | 21.48 |
| 1 | 2.0 | 20.24 |
| 1 | 2.5 | 19.12 |
| 1 | 3.0 | 18.33 |
| 1 | 3.5 | 11.01 |
| 1 | 4.0 | −5.82 |
| 1 | 4.1 | −6.25 |

**الجدول (2) يبين قيم SDR عند تغيير نسبة مطالي الإشارتين**

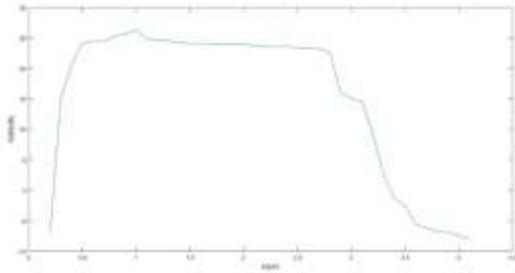

**الشكل-3- نسبة الإشارة للتشوه بدلالة نسبة مطالي الإشارتين الممزوجتين**

نلاحظ من الشكل -3- أن أداء الخوارزمية يكون أفضل كلما كان مطالي الإشارتين أقرب لبعضهما (على عكس حالة الترددات)، وإذا أردنا تحديد مجال عمل الخوارزمية فتكون نسبة الإشارة للتشوه أعلى من 15dB طالما أن نسبة مطالي الإشارتين تنتمي للمجال [0.3,3]، أي مطال إحداهما ثلاثة

أضعاف الأخرى أو أقل، وينحدر الأداء بشدة عندما تكون نسبة المطالات خارج المجال السابق.

نستنتج من الدراسة التي تمّت في الفقرة (4-1) و (4-2) وجود شرطين لتستطيع خوارزمية CEEMDAN أن تفصل بين الإشارات الممزوجة هما:

1- أن تكون نسبة ترددي الإشارتين أقل من 0.6 أو أكبر من 1.6.
2- أن تكون نسبة مطالي الإشارتين ضمن المجال [0.3,3].

### 4-3- دراسة أداء الخوارزمية CEEMDAN لفصل المنابع الصوتية

تعالج التجارب التالية حالة فصل الضجيج عن الكلام بواسطة الخوارزمية CEEMDAN، وفصل إشارتين كلاميتين عن بعضهما. تمّ في كل مرحلة القيام بعدة تجارب لدراسة إمكانيات وحدود الخوارزمية، وأُستخدمت بيئة الماتلاب لإجراء التمثيل الحاسوبي.

في المرحلة الأولى تمّ الاستفادة من قاعدة معطيات Noizeus[1] والتي تحوي إشارات كلامية مضججة بقيم مختلفة من نسبة الإشارة للضجيج SNR، طُبق على هذه الإشارات خوارزمية CEEMDAN وجرى مناقشة الخرج الناتج. مُزجت اشارتان كلاميتان في المرحلة الثانية وطُبق المزيج على دخل الخوارزمية المذكورة، فأظهرت النتائج عجز الخوارزمية عن فصل الاشارات الكلامية عن بعضها.

### 4-3-1- فصل الضجيج عن الكلام

تبرز أهمية مسألة فصل الضجيج عن الكلام في التطبيقات التي تحتاج لدراسة الضجيج بمعزل عن الكلام والذي يعتبر في هذه الحالة إشارة تداخل غير مرغوبة، مثل الرغبة في مراقبة صوت آلة معينة لمعرفة حالتها الفنية وحذف أصوات العمال من حولها، واستخدام CEEMDAN لفصل الضجيج عن الكلام قد يعطي كلتا الإشارتين (الكلام والضجيج)، وهذا ما يميز طريقة CEEMDAN عن طرق تحسين الكلام الأخرى والتي تعطي الإشارة الكلامية المحسَّنة فقط.

باستخدام بيئة ماتلاب للمحاكاة واختيار نوع الضجيج (Bubble, Airport, Car) وقيم مختلفة لنسبة الإشارة للضجيج SNR = {0,5,10,15} dB تمّ الحصول على النتائج المبينة في الجدول(3):

| Noise | Babble | | | | Airport | | | | Car | | | |
|---|---|---|---|---|---|---|---|---|---|---|---|---|
| SNR | 0dB | 5dB | 10dB | 15dB | 0dB | 5dB | 10dB | 15dB | 0dB | 5dB | 10dB | 15dB |
| SDR(dB) | -1.00 | 3.03 | 6.65 | 8.63 | -1.53 | 2.32 | 7.12 | 9.25 | -3.17 | 2.17 | 4.53 | 8.62 |
| SAR(dB) | -0.89 | 3.07 | 6.70 | 8.58 | -1.44 | 2.35 | 7.21 | 9.30 | -3.04 | 2.25 | 4.57 | 8.68 |
| PESQ | 1.81 | 2.03 | 2.25 | 2.38 | 1.65 | 1.97 | 2.25 | 2.46 | 1.72 | 1.90 | 2.05 | 2.30 |

**الجدول (3) نتائج فصل بعض أنواع الضجيج عن الكلام**

نلاحظ أن أداء الخوارزمية جيد جداً في قيم SNR العالية ولكنه ينخفض كثيراً في القيم الدنيا. رسمنا كل من الإشارة المضججة والنقيّة والمحسَّنة في الطيف Spectrogram والزمن في حالة نوع الضجيج Bubble عند قيمة SNR = 15dB فكانت النتائج كما في الشكلين 4 و5.

---

[1] https://ecs.utdallas.edu/loizou/speech/noizeus/

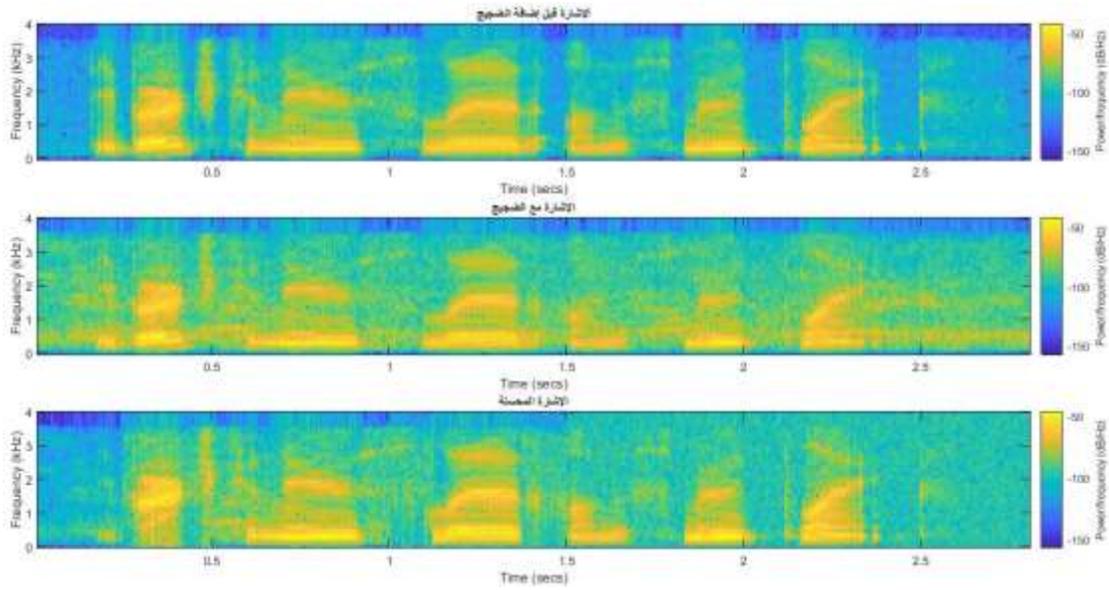

**الشكل-4- طيف كل من الاشارات النقية والمضججة والمحسنة عند قيمة SNR=15dB**

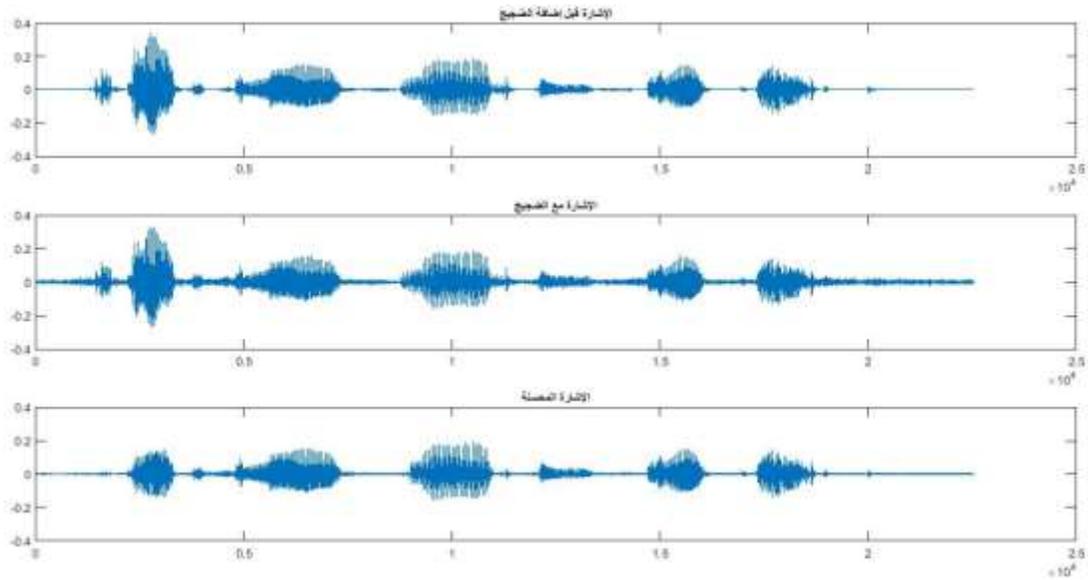

**الشكل-5- الاشارات النقية والمضججة والمحسنة في المجال الزمني عند SNR=15db**

### 4-3-2- فصل اشارة كلامية عن اشارة كلامية أخرى

يُلاحظ لدى مزج اشارتين كلاميتين مزجاً لحظياً وتطبيق خوارزمية CEEMDAN عدم تشابه النمطين الأول والثاني مع الإشارتين قبل المزج (من الشكل-6- أو بالاستماع)، وبالتالي لم تتمكن خوارزمية CEEMDAN من فصل الإشارتين عن بعضهما، وهذا منطقي لأن الإشارات الكلامية متداخلة في التردد والمطال حتى لو اختلف المتكلمين، و لم يتحقق شرطَي عمل طريقة CEEMDAN.

يتضح من الشكلين 4 و 5 قدرة الخوارزمية على حذف الضجيج ويمكن ملاحظة ذلك بوضوح بجوار العينة رقم $2 \times 10^4$، كما نلاحظ بعض التشوه في مطال الإشارة المحسَّنة ويظهر هذا التشوه جلياً بالقرب من العينة رقم $0.25 \times 10^4$ ، وهذا سببه تكرار عمليات الطرح خلال الخوارزمية حيث أن مجموع الأنماط الناتجة يجب أن يساوي الإشارة الأصلية، وبالتالي فإن قيمة مطال أي نمط عند عينة ما يجب أن يكون أصغر من قيمة مطال الإشارة الأصلية عند تلك العينة.

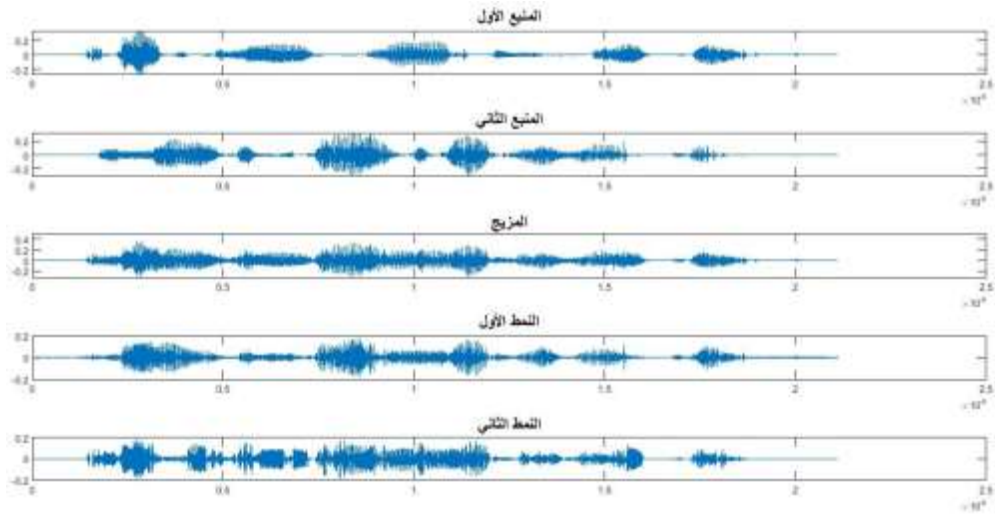

الشكل-6- نتائج تطبيق CEEMDAN على مزيج من اشارتين كلاميتين حيث يظهر الإشارتين قبل المزج والمزيج ونتائج الفصل

بمقارنة نتائج خوارزمية CEEMDAN في فصل الضجيج عن الكلام مع طريقة SEGAN وWIENER (Pascual Santiago, et al., 2017) حسب المعيار PESQ الذي يعبّر عن وضوح الكلام وجودته فكان أداء خوارزمية CEEMDAN قريباً من أداء تلك الطريقتين وفق هذا المعيار كما في الجدول (4).

| Method | WIENER | SEGAN | CEEMDAN |
|---|---|---|---|
| PESQ | 2.22 | 2.16 | 2.15 |

**الجدول (4)** مقارنة بين **CEEMDAN** وكل من طريقة **SEGAN** و**WIENER** حسب المعيار **PESQ**

تمّ إيجاد النتيجة في الجدول (4) بتطبيق الخوارزمية على نفس قاعدة المعطيات التي طُبق عليها SEGAN وWIENER وهي Voice Bank Corpus (Christophe, et al., 2013). نستنتج من الجدول (4) أن CEEMDAN أعطت نتائج أقل بقليل من الطريقتين SEGAN وWIENER ولكن تبقى جيدة لأنها لا تحتاج لتدريب شبكة عصبونية كما في SEGAN ولا إلى تقدير بعض المعاملات كما في WIENER.

إنّ خوارزمية CEEMDAN لا تحتاج إلى شروط معينة ويمكن تطبيقها على كافة الإشارات فهي مقادة بالمعطيات Data-driven Method وهي تتميز عن طرق تحسين الكلام أنها لا تعطي الإشارة المحسَّنة فقط بل تعطي إشارة الضجيج أيضاً والتي قد تكون صوت لآلة معينة لتحديد حالتها الفنية.

من مساوئ خوارزمية CEEMDAN أنها تحلل الإشارة إلى عدد غير ثابت من الأنماط وهي المشكلة نفسها التي عانت منها EMD فلم تستطع CEEMDAN حلّها مما يجعلها غير قادرة على العمل على نحو متدفق Streaming. يُضاف إلى ذلك أيضاً التشوه المطالي في الإشارة المحسَّنة بسبب تكرار عمليات طرح الأنماط.

## 5- خاتمة

تمّ في هذا البحث إلقاء الضوء على خوارزمية CEEMDAN الحديثة نسبياً، واستنتاج شروط عمل هذه الخوارزمية بتطبيقها على إشارتين جيبيتين، وجرى تقييم أدائها في فصل المنابع الصوتية، ويظهر جلياً قدرتها على حذف بعض أنواع الضجيج وذلك عندما تكون نسبة الترددات ونسبة المطالات تحقق شرطَيْ عملها، ولكنها غير مناسبة لفصل الإشارات الكلامية عن بعضها. لا تحتاج هذه الخوارزمية الانتقال إلى المجال الترددي، فقط عمليات تكرارية في المجال الزمني كافية لتحليل الإشارة إلى الأنماط الأساسية فيكون أحد الأنماط هو الإشارة المحسَّنة والآخر هو الضجيج، كما أنها لا تتطلب شروط على الاشارة المدروسة Data-driven Method. من الآفاق المستقبلية لهذه الخوارزمية أنه من الممكن أن تتكامل مع أنظمة أخرى لزيادة دقة فصل الضجيج عن الكلام، واستخدام طرق التعلّم العميق في فصل الإشارات الكلامية.

## المراجع